\documentclass[prl,twocolumn,showpacs]{revtex4}
\usepackage[xdvi]{graphicx}%
\usepackage{dcolumn}
\usepackage{amsmath}
\usepackage{color}
\usepackage{ulem}

\newcommand{\bv}[1]{{\boldsymbol #1}}

\begin{document}

\title{
Computation of Large Deviation Statistics via Iterative
Measurement-and-Feedback Procedure}

\author{Takahiro Nemoto and Shin-ichi Sasa} 

\affiliation{Division of Physics and Astronomy, Graduate School of Science,
Kyoto University, Kyoto 606-8502, Japan}
\date{\today}

\begin{abstract}
We propose a computational method for large deviation statistics 
of time-averaged quantities in general Markov processes. 
In our proposed method, 
we repeat a response measurement
against external forces, where the forces are determined by the previous measurement as
feedback.
Consequently, we obtain a set of stationary 
states corresponding to an
{\it exponential family} of distributions, 
each of which shows rare events in 
the original system as the typical behavior. 
As a demonstration of our method,
we study large deviation statistics of one-dimensional lattice gas models.
\end{abstract}


\maketitle



{\it Introduction.}---Rare events, which are hardly observed, sometimes lead to substantial 
effects in nature. Examples may be seen in a broad range of 
phenomena including biomolecular reactions, nucleation, and plate-tectonic activities. Although understanding the statistical behavior of such rare 
events may be an important problem, it turns out from the 
definition that direct observation of rare events 
is too difficult.
Remarkably, several 
techniques have been invented for 
generating rare events in numerical simulations,
such as transition path 
sampling \cite{Frenkel, Chandler}, 
transition interface sampling \cite{Erp,Erp2}, 
forward flux sampling \cite{Allen,Allen2}, 
and the population dynamics method \cite{Populationdynamics,Populationdynamics2}. 
However, contrary to the progress made with respect to numerical simulations, 
there are no methods that facilitate
the observation of rare events in laboratory systems. 
Our final goal is to construct
a rare-event sampling method that can be useful in laboratory 
experiments. Toward this end, in the present Letter,
for rare events characterized by {\it large deviation statistics}, 
we efficiently calculate the frequencies of the rare events 
via an iterative measurement-and-feedback procedure, which might be implemented in laboratory
experiments.


We here formulate the large deviation statistics.
For a time-dependent quantity $x(t)$, 
we consider the time-averaged value $X(\tau)=(1/\tau)\int_{0}^{\tau} dt x(t)$,
where $\tau$ denotes the averaging time. According to 
the law of large numbers, as we increase $\tau$, 
the deviation of $X(\tau)$ from the expected value $X_0$
decreases. However, even for large values of $\tau$, when we perform 
the same measurement many times, certain obtained values will 
deviate from the typical value $X_0$. 
The large deviation principle, which is proved in many systems \cite{Dembo_Zeitouni,Touchette}, is a law that claims 
the asymptotic form of the probability density in the limit $\tau \rightarrow \infty$ to be $e^{-\tau I(X)}$. Here, the function
$I(X)$, which characterizes the statistical properties of those rare events,  is called a large deviation function.



While studies of large deviation functions have a long history
in probability theory, the functions have  
attracted attention in the field of nonequilibrium 
physics particularly in the last two decades.  Since the discovery
of the fluctuation theorem, which is the symmetry
property of the large deviation function of the
time-averaged entropy production rate \cite{FT,FT2}, 
several results for large deviation functions have been found, such as
an additivity principle for driven diffusive systems \cite{BD,BD2}, 
dynamical phase transitions of kinetically constrained models
\cite{garrahan,garrahan2}, 
a Lyapunov function for 
nonequilibrium steady states without relying on entropy production \cite{CMaes},
exact results for the current statistics of 
lattice gas models \cite{Gorissen,Gorissen2}, 
and formulas motivated by the formal correspondence 
between large deviation functions and thermodynamic functions \cite{Evans,JackSollich,
Discussion_of_modifiedTransitionrate,Chetrite_Touchette}.
All of this progress clearly indicates
that large deviation statistics plays a key role 
in nonequilibrium physics. 
Therefore, our rare-event sampling formulation
will shed light 
on the development of nonequilibrium statistical mechanics,
along with the long-term goal stated 
in the first paragraph.


The computational method we propose in this Letter consists
of the iteration of a response measurement against 
external forces, where the measurement result 
determines the next external forces we add for the next measurement.
By this method, we obtain a set of stationary states corresponding to an
{\it exponential family} of path probability densities, from which the large deviation statistics
may be constructed \cite{Dembo_Zeitouni,Touchette}. 
We apply our computational method to one-dimensional lattice gas models.
As a result, we present some suggestions about rare fluctuations of 
those models.

{\it Set up.}---Let the state space $\Omega$ be a finite set. 
We consider continuous time Markov processes on the space 
$\Omega$. For $\bv n, \bv n' \in \Omega$, we define 
a transition rate $w(\bv n \rightarrow \bv n^{\prime})$. The 
escape rate is then determined as $\lambda(\bv n) \equiv 
\sum_{\bv n^{\prime}\in \Omega }w(\bv n\rightarrow \bv n^{\prime})$. 
The transition rate is assumed to be an irreducible matrix 
that satisfies $w(\bv n\rightarrow \bv n)=0$ and 
$w(\bv n\rightarrow \bv n^{\prime})\neq 0$ if 
$w(\bv n^{\prime}\rightarrow \bv n)\neq 0$ . 
The history of states during a time interval $\tau$, which is denoted by $\omega$,
is specified by the total number of transitions $n$, a collection of 
transition times $(t_i)_{i=1}^{n}$, and a sequence of 
states $(\bv n_i)_{i=0}^{n}$, 
where $\bv n_{i}=\bv n(t)$ for $t_{i}\leq t \leq t_{i+1}$ with
$t_0 \equiv 0$, $t_{n+1}\equiv \tau$. We denote 
the path probability density in the steady state 
by $P(\omega)$, and the expected value with respect to  $P(\omega)$
is represented by $\left \langle \ \right \rangle$.


We study the statistical properties of a quantity 
$\alpha(\bv n_i \rightarrow \bv n_{i+1})$ defined 
at each transition $\bv n_i \rightarrow \bv n_{i+1}$ 
and its time-averaged value 
$A(\omega) = (1 / \tau)\sum_{i=0}^{n-1}\alpha(\bv n_i \rightarrow \bv n_{i+1})$.
The large deviation principle of $A(\omega)$ is that the probability 
density of $A(\omega)$ obeys $p(A) \simeq e^{-\tau I(A)}$ for large values of $\tau$, 
where $I(A)$ denotes the large deviation function. Here, we introduce 
a dynamical free energy (or scaled cumulant generating function) 
defined by
\begin{equation}
G(h)
\equiv \lim_{\tau\rightarrow \infty}\frac{1}{\tau} 
\log \left \langle e^{h\tau A(\omega)}\right \rangle,
\label{defG(h)}
\end{equation}
where $h$ is called a biasing parameter. 
Similar to thermodynamics, the dynamical free energy 
is equivalent to the large deviation function through the 
Legendre transformation  $I(A)=\max_{h}[hA - G(h)]
$\cite{Dembo_Zeitouni, Touchette}.  Furthermore, the dynamical free energy
is closely related to the {\it exponential family} defined by 
$P(\omega;h)=P(\omega) e^{h \tau A(\omega)}/\left \langle 
e^{h\tau A(\omega)}\right \rangle$. (Indeed, direct calculation shows
that the expected value of $A(\omega)$ with respect to $P(\omega;h)$ in
the limit $\tau \rightarrow \infty$
is equal to 
$\partial G(h)/\partial h$.) Thus, hereafter, we focus on the exponential
family instead of the large deviation function $I(A)$. 


We note that 
the most dominant contribution to 
$\left \langle e^{h\tau A(\omega)}\right \rangle$ for large values of $\tau$ comes from trajectories  
that satisfy $\partial I(A)/\partial A|_{A=A(\omega)}=h$.
This means that rare trajectories are necessary to evaluate 
$\left \langle e^{h\tau A(\omega)}\right \rangle$, and thus
evaluation with respect to the exponential family is 
quite hard. The problem we solve is to find an 
efficient method to calculate the expected values in the exponential family.

{\it Evolution in an exponential family.}---A basic strategy of our computational method for large deviation 
statistics is to construct stationary states corresponding to the exponential family along the
$h$ axis by iterating a measurement and feedback. The procedure
is as follows. We fix a measurement time $\tau$ that is much larger
than the correlation time of $\alpha$, $\tau_\alpha$, 
and we choose a small increment 
$\delta h$ such that $ \tau \delta h \left \langle A(\omega) \right 
\rangle =O(1)$. As a first step,  
we measure  
$\left \langle e^{\tau \delta h   A(\omega)} \right \rangle_{\bv n}$
as a function of $\bv n$ in the original system, 
where $\left \langle \ \right \rangle_{\bv n}$ represents 
the expected value with the initial condition $\bv n(0)=\bv n$ fixed. 
Here, we note that the measurement is not hard to perform due to the condition $\tau \delta h \left \langle A(\omega)\right \rangle=O(1)$.
Subsequently, depending on the value of 
$\left \langle e^{\tau \delta h A(\tau)} \right \rangle_{\bv n}$,
we modify the transition rate as 
\begin{equation}
w^{\delta h}(\bv n\rightarrow \bv n^{\prime}) = w(\bv n\rightarrow \bv n^{\prime}) e^{\delta h \alpha(\bv n \rightarrow \bv n^{\prime})}\frac{\left \langle e^{ \tau \delta h A(\omega)} \right \rangle_{\bv n^{\prime}}}{\left \langle e^{\tau \delta h A(\omega)} \right \rangle_{\bv n}}.
\label{wh_1}
\end{equation}
Next, in this modified system, we measure the expected value 
of the same quantity $e^{\tau \delta h A(\omega)}$, which we denote by 
$\left \langle e^{ \tau \delta h A(\omega)} \right \rangle_{\bv n}^{\delta h}$.
We then define the second modified transition rate
\begin{equation}
\begin{split}
&w^{2\delta h}(\bv n\rightarrow \bv n^{\prime}) = w^{\delta h}(\bv n\rightarrow \bv n^{\prime}) e^{\delta h \alpha(\bv n \rightarrow \bv n^{\prime})}\frac{\left \langle e^{ \tau \delta h A(\omega)} \right \rangle_{\bv n^{\prime}}^{\delta h}}{\left \langle e^{ \tau \delta h A(\omega)} \right \rangle_{\bv n}^{\delta h}}.
\end{split}
\end{equation}
By iterating this procedure, we obtain a set of transition rates
\begin{equation}
\begin{split}
&w^{l \delta h}(\bv n\rightarrow \bv n^{\prime}) = w(\bv n\rightarrow \bv n^{\prime}) e^{l \delta h \alpha(\bv n \rightarrow \bv n^{\prime})}\prod_{k=0}^{l-1}\frac{\left \langle e^{ \tau \delta h A(\omega)} \right \rangle_{\bv n^{\prime}}^{k \delta h}}{\left \langle e^{ \tau \delta h A(\omega)} \right \rangle_{\bv n}^{k \delta h}}
\label{whmesured}
\end{split}
\end{equation}
with $l=0,1,2, ...$.
Our computational method is based on the following formula.
Let $\left \langle f \right \rangle^{h}$ be 
the expected value of time-extensive quantities $f(\omega)$ 
in the system with the modified transition rate $w^h$  
($h=0,\delta h, 2\delta h, ...$). We then have 
\begin{equation}
\left \langle f(\omega) \right \rangle^{h} 
\simeq \frac{\left \langle f(\omega) e^{h\tau A(\omega)} 
\right \rangle}{\left \langle e^{h\tau A(\omega)} \right \rangle}.
\label{mathBasis}
\end{equation}
Here and hereafter, $\simeq$ represents the asymptotic equality 
when  $\tau \gg \tau_\alpha$.


We show the outline of the derivation of (\ref{mathBasis}).
For a given $u^{h^{\prime}}(\bv n^{\prime} \rightarrow \bv n)$, 
we define a matrix
\begin{equation}
L^{h,h^{\prime}}_{\bv n,\bv n^{\prime}}\equiv u^{h^{\prime}}(\bv n^{\prime}\rightarrow \bv n) e^{h\alpha(\bv n^{\prime}
\rightarrow \bv n)} - \lambda^{u^{h^{\prime}}}(\bv n) \delta_{\bv n, \bv n^{\prime}},
\label{matrixL}
\end{equation}
where $\lambda^{u^{h^{\prime}}}(\bv n) \equiv \sum_{\bv n^{\prime}} u^{h^{\prime}}(\bv n\rightarrow\bv n^{\prime})$.
Let $\phi^{h,h^{\prime}}$ be the left eigenvector
of the largest eigenvalue problem of (\ref{matrixL}). 
Subsequently, we construct $u^{h^{\prime}}(\bv n\rightarrow \bv n^{\prime})$ from $w(\bv n\rightarrow \bv n^{\prime})$ as follows.
First, we set $u^{h^{\prime}}(\bv n\rightarrow \bv n^{\prime})|_{h^{\prime}=0}\equiv w(\bv n\rightarrow \bv n^{\prime})$, from which $\phi^{h,0}$
is determined. (The corresponding eigenvalue is known to be equal to $G(h)$ \cite{garrahan2,Touchette}.) Next, from $\phi^{h,0}$, we define 
$u^{h}(\bv n^{\prime}\rightarrow \bv n) \equiv w(\bv n^{\prime}\rightarrow \bv n) e^{h \alpha(\bv n^{\prime} \rightarrow \bv n)} \phi^{h,0}(\bv n)/\phi^{h,0}(\bv n^{\prime})$.
Now, for this eigenvalue problem, we can prove {\it the multiplicative property} 
\begin{equation}
\phi^{h+h^{\prime},0}= \phi^{h,h^{\prime}} \phi^{h^{\prime},0}.
\label{key1}
\end{equation}
This is a key relation to derive (\ref{mathBasis}).
Indeed, 
by using the standard technique for the cumulant generating function (see, for example, Ref. \cite{garrahan2}), we have
$\phi^{\delta h, 0}(\bv n) \simeq \left \langle e^{\tau \delta h A(\omega)} \right \rangle_{\bv n}$, which leads to $w^{\delta h}\simeq u^{\delta h}$. By applying the same relation to the modified system $u^{\delta h}$, we obtain
$\phi^{2\delta h, \delta h}(\bv n) \simeq \left \langle e^{\tau \delta h A(\omega)} \right \rangle_{\bv n}^{\delta h}$. 
Subsequently, using (\ref{key1}), we can show $\phi^{2\delta h,0}(\bv n)\simeq \prod_{k=0}^{1}\left \langle e^{\tau \delta h A(\omega)} \right \rangle_{\bv n}^{k\delta h}$ leading to $w^{2\delta h} \simeq u^{2\delta h}$. 
After iterating this procedure $M$ times, we eventually obtain
\begin{equation}
\phi^{l \delta h,0}(\bv n)\simeq \prod_{k=0}^{l-1}\left \langle e^{\tau \delta h A(\omega)} \right \rangle_{\bv n}^{k\delta h},
\label{mathemtaicalclaim}
\end{equation}
which leads to $w^{l\delta h}\simeq u^{l\delta h}$ for $l=1,2,...$.
Since the expected values of time-extensive quantities in the steady state generated by $u^h$
are nearly equal to the same quantities with respect to the distribution of the exponential family \cite{JackSollich,Discussion_of_modifiedTransitionrate,Chetrite_Touchette},
we arrive at (\ref{mathBasis}).
See \cite{suppl1} for the details of the derivation.


{\it Effective description of the exponential family.}---Up to this point, we have confirmed that the rare trajectories most contributing 
to $G(h)$ are generated by the modified transition rate 
$w^h(\bv n \to \bv n')$ in (\ref{whmesured}). Here, we consider applications
of the formula to spatially extended many-body systems such as lattice gas 
models consisting of $L$ sites. Consequently, the computation time for obtaining
$w^h(\bv n \to \bv n')$ is proportional to $O(2^L)$ in the large $L$
limit, because  we need to measure 
$\left \langle e^{\tau \delta h A(\omega)} \right \rangle_{\bv n}^{l\delta h}$ 
for each $\bv n$. This means that the application to many-body systems 
is still hard. Nevertheless, 
as seen in many examples in statistical
physics \cite{Oonobook}, we expect that there might be a simple model that describes an 
essential feature of rare-event phenomena. Along with this expectation, 
we introduce {\it an effective description of the exponential family},
where we assume that the details of the modified transition rate do not 
seriously affect the statistical properties of rare fluctuations.
More precisely, we introduce an effective transition rate with $K$ unknown parameters,
where $K\ll O(2^L)$ for each value of $h$.
Then, by employing (\ref{whmesured}), we determine the values of those parameters.

\begin{figure}[tbh]
\includegraphics[width=4cm]{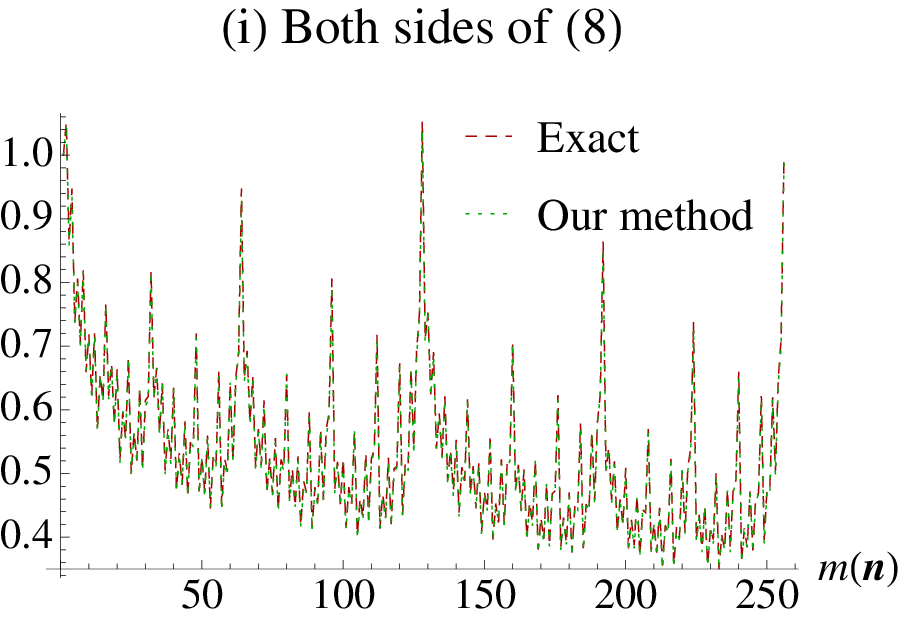}
\includegraphics[width=4cm]{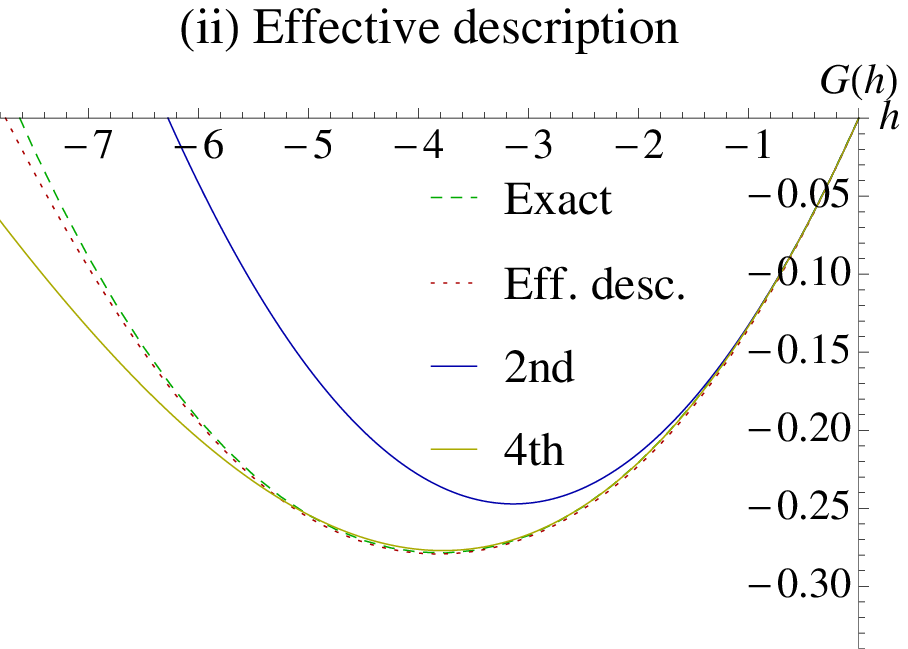}
\caption{(color online) Statistical properties of current in an open boundary ASEP obtained by our computational method. We set $q=0.5$, $L=8$, $\alpha=0.8$, $\beta=0.8$, $\gamma=0.2$, and $\delta=0.2$. (i) We plot the right-hand side of (\ref{mathemtaicalclaim})
for $l=100$ and $\delta h=-0.02$ obtained from Monte Carlo simulations with (\ref{whmesured}) (Our method). We also plot the left-hand side
of (\ref{mathemtaicalclaim}) obtained by the diagonalization
of the matrix (\ref{matrixL}) for $l \delta h = -2$ (Exact). The $x$ axis 
represents $m(\bv n)=\sum_{i=0}^{L-1}n_{L-i}2^i$,
which is the decimal value of the binary number $\bv n$. These two lines coincide with each other.
(ii) $G(h)$ obtained from Monte Carlo simulations with the effective modified transition rate (\ref{effectiveTransitionASEP}) (Eff. desc.). We also plot the largest eigenvalue of (\ref{matrixL}) with $h^{\prime}=0$ (Exact). Furthermore,
we plot the truncated cumulant expansions up to the second order (2nd) and the fourth order (4th), which were calculated using the exact formula of ASEP \cite{Gorissen,Gorissen2}.
}
\label{ASEP_Exactpotential}
\end{figure}


{\it Asymmetric simple exclusion process (ASEP).}---As a demonstration of our method,
we study the large deviation statistics of one-dimensional lattice gas models.
We first consider a lattice of size $L$ with open boundary 
conditions, where each of the sites 
accommodates at most one particle. We denote a particle configuration by $\bv n\equiv (n_i)_{i=1}^{L}$, where $n_i$ 
takes a value of $1$ (occupied) or $0$ (empty). The transition rate 
$w(\bv n\rightarrow \bv n^{\prime})$ is defined as follows.
A particle moves to the left empty site
with a rate $q$ and to the right empty site with a rate $1$. 
A particle is injected into the boundary site $i=1$ ($i=L$) at a rate
$\alpha$ ($\delta$) and the particle at the boundary site $i=1$ ($i=L$) 
is removed at a rate $\gamma$ ($\beta$). This model is called ASEP.
We focus on bulk current defined as
$\alpha (\bv n\rightarrow \bv n^{\prime})
=1/(L-1)\sum_{i=1}^{L-1} j_{i}(\bv n\rightarrow \bv n^{\prime})$,
where $j_{i}(\bv n\rightarrow \bv n^{\prime})$ takes the value of $1$ (or $-1$) 
when a particle  moves from $i$ to $i+1$ ($i+1$ to $i$).
By evaluating the transition rate (\ref{whmesured}) by Monte Carlo simulations
and comparing it with the exact result obtained from the diagonalization of the matrix (\ref{matrixL}), we numerically confirm our formula (\ref{mathBasis}).
See Fig. \ref{ASEP_Exactpotential} (i)
for an example of the obtained results.

Next, we study an effective description of the exponential family. 
We introduce an effective transition rate $w_{\rm eff}^{h}$ with $L+1$ unknown parameters
$( \psi_{h,i} )_{i=0}^{L}$. Concretely, we define
\begin{equation}
w_{\rm eff}^h(\bv n\rightarrow C_{i\rightarrow i\pm 1}\bv n)\equiv w(\bv n\rightarrow  C_{i\rightarrow i\pm1}\bv n)e^{\pm h/(L-1)}\frac{\psi_{h,i\pm1}}{\psi_{h,i}},
\label{effectiveTransitionASEP}
\end{equation}
where $C_{i\rightarrow i\pm 1}$ denotes an operator that moves a particle from the site $i$ to the site $i\pm 1$.
For the left (or right) boundary transition, we also define $w_{\rm eff}^h(\bv n\rightarrow C_{\rm{L},\pm 1}\bv n)\equiv w(\bv n\rightarrow  C_{\rm{L},\pm 1}\bv n) (\psi_{h,1}/\psi_{h,0})^{\pm 1}$ (or $w_{\rm eff}^h(\bv n\rightarrow C_{\rm{R},\pm 1}\bv n)\equiv w(\bv n\rightarrow  C_{\rm{R},\pm 1}\bv n) (\psi_{h,L}/\psi_{h,0})^{\pm 1}$), where $C_{{\rm L \ or \ R},a}$ denotes an operator that injects a particle into the boundary site (left or right) when $a=+1$ and removes a particle from the boundary site (left or right) when $a=-1$.
The effective transition rate corresponds to an ASEP that has a new spatially varying 
transition rate. 
In other words, the effective transition rate is given by adding a one-body potential to the system.


The values of the parameters $( \psi_{h,i} )_{i=0}^{L}$ are determined by 
our computational method.
Let us suppose that we already have the values of the parameters $( \psi_{l\delta h,i} )_{i=0}^{L}$. Then, we measure $ \left \langle e^{\tau \delta h A(\omega)}
\right \rangle_{\bv n}^{l\delta h}$ for $L+1$ different configurations 
$\bv{n}=\bv{n}_j$ ($j=0,1, 2, ...,  L)$. In particular, here, we set $(\bv{n_j})_{i}=\delta_{ij}$ as the simplest choice.
Next, by applying (\ref{whmesured}) to the effective transition rate (\ref{effectiveTransitionASEP}), we obtain the following $L+1$ equations that connect the next parameters 
with
the previous ones and 
the observed quantities:
$\psi_{(l+1)\delta h,i} = \psi_{l\delta h,i} \left \langle e^{\tau \delta h A(\omega)}
\right \rangle_{\bv n_i}^{l\delta h}$ ($i=0,1,2,...,L$).
In this manner, the computation time becomes proportional to $O(L)$,
which is substantially reduced from $O(2^L)$.


By employing our computational method, we calculated $G(h)$, which is plotted 
in Fig. \ref{ASEP_Exactpotential} (ii). In the same figure, for comparison, 
we also plot the exact result obtained from the eigenvalue problem of (\ref{matrixL})
and the truncated cumulant expansions up to the second and 
the fourth order 
obtained from the exact formula in Refs. \cite{Gorissen,Gorissen2}.
Although there is a small deviation between our result (red dotted line) 
and the exact result (green dashed line) around $h=-7$, the accuracy of our result
is considerably better than the result for the truncated cumulant expansions
(blue and yellow solid lines). This result indicates 
that rare fluctuations of the ASEP with this parameter set 
are well characterized by the effective transition 
rate (\ref{effectiveTransitionASEP}).
We expect that there exists a mathematical formula related to this observation.
We remark that the variational expression proposed in
\cite{BD,BD2} can be derived from another variational principle provided in Refs.
\cite{Discussion_of_modifiedTransitionrate,Discussion_of_modifiedTransitionrate2} if we are allowed
to use the effective transition rate (\ref{effectiveTransitionASEP}) in the limit $L\rightarrow \infty$.
We will study the effective transition rate (\ref{effectiveTransitionASEP}) more systematically in future.

\begin{figure}[tbh]
\includegraphics[width=4.2cm]{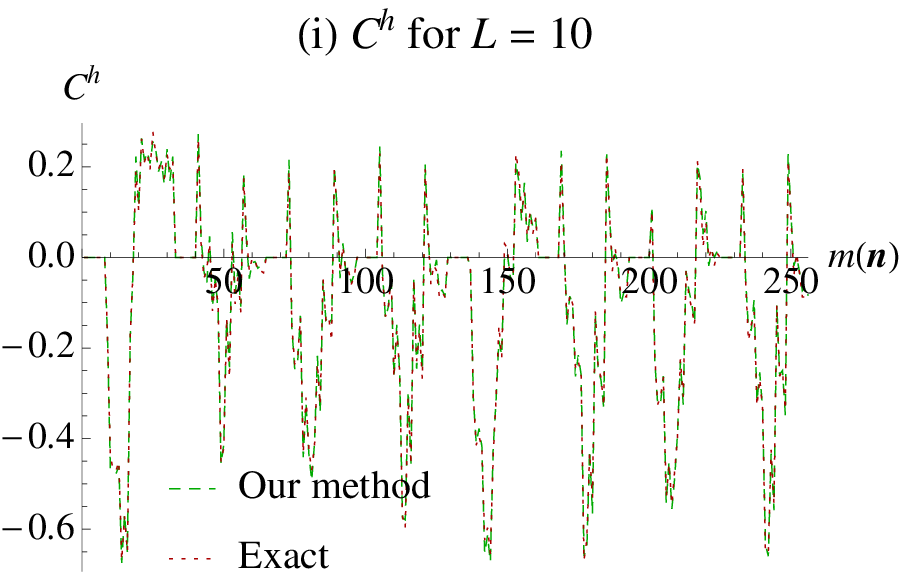}
\includegraphics[width=4cm]{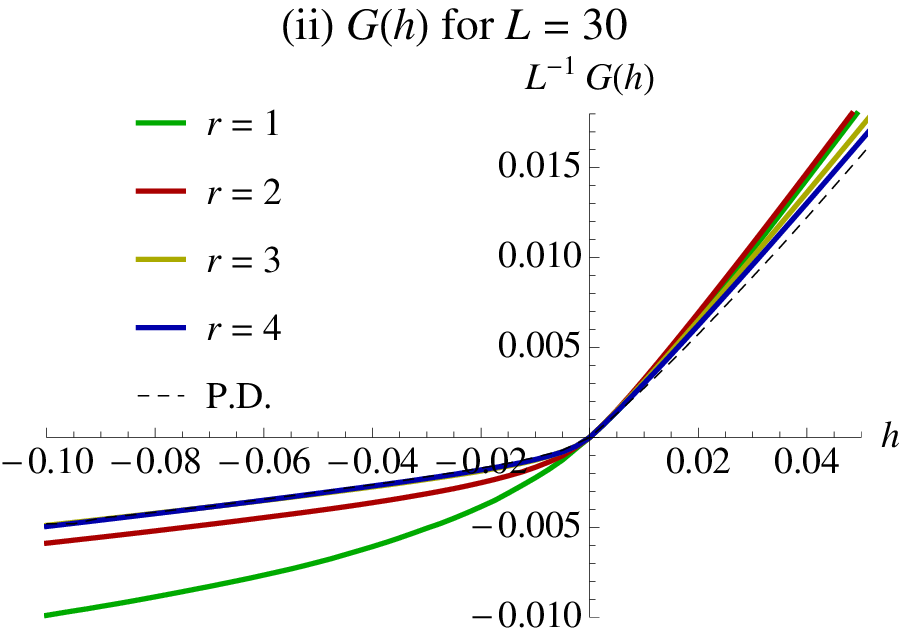}
\caption{(color online)
Statistical properties of an activity in a FA model for $c=0.3$.
(i) For $L=10$ with $r=4$, we plot $C^{l \delta h}((n_{1+j})_{j=1}^{r})$ obtained from our method (Our method), where we set $l=40$ and $\delta h=-0.0025$. The $x$ axis represents $m(\bv n)=\sum_{i=0}^{r-1}n_{L-i}2^i+\sum_{i=r}^{2r-1}n_{2r+1-i}2^i$. On the same figure, we also plot
$\phi^{l \delta h,0}(F_1 \bv n)/\phi^{l \delta h,0}(\bv n)$ with $\bv n = (0,n_2,...,n_{r+1}, 0, ... , 0, n_{L-r+1},... n_{L} )$ obtained from the  diagonalization of the matrix (\ref{matrixL}) for $L=10$, $l \delta h  =-0.1$ (Exact).
These two lines coincide each other.
(ii) $G(h)/L$ as a function of $h$ with the truncating 
number $r=1,2,3$, and $4$ for $L=30$.
We also plot the result obtained from the population 
dynamics (P. D.) method \cite{Populationdynamics,Populationdynamics2} 
as a black dashed line. 
} 
\label{fig_modification_of_fluctuation_theorem}
\end{figure}


{\it Fredrickson--Andersen (FA) model.}---We next consider a FA model in a one-dimensional lattice 
of size $L$ with periodic boundary conditions.  
An occupation variable $n_i$, which takes a value of
0 (empty) or 1 (occupied),
is defined for each site $i$. From a configuration $\bv n=(n_i)_{i=1}^L$ to  $F_i\bv n \equiv (n_1,n_2,...,1-n_i,...,n_L)$, 
a transition rate is defined as $w(\bv n \rightarrow F_i\bv n) \equiv [(1-c)n_i+c(1-n_i)]f_i(\bv n)$, where the function 
$f_i(\bv n)\equiv n_{i-1}+n_{i+1}$ represents the kinetic constraints.
Since the detailed balance condition is independent of $f_i(\bv n)$,
the stationary probability $p(\bv{n})$ is derived as 
$p(\bv{n})=\prod_{i=1}^L [c n_i+(1-c)(1-n_i)]$.
Although the stationary state is trivial, the system exhibits a
dynamical phase transition \cite{garrahan,garrahan2}, which might
be related to
dynamical heterogeneities \cite{KCMReview1,KCMReview2}.
In order to study the dynamical phase transition, we consider the large deviation statistics of the time-averaged value of an
activity defined as $\alpha(\bv n\rightarrow \bv n^{\prime})=1$.
The dynamical 
phase transition has been determined
as the singularity of the cumulant 
generating function in the limit $L \to \infty$ \cite{garrahan,garrahan2}.


Here, we apply our method to this model. 
We define the effective transition rate as
$w^h_{\rm eff}(\bv n\rightarrow  F_i \bv n)\equiv w(\bv n\rightarrow  F_i \bv n)e^{h}[C^h((n_{i\pm j})_{j=1}^{r})]^{1-2n_i}$, where $r$ is a truncating number of the interaction range and $C^h((n_{i\pm j})_{j=1}^{r})$
is an unknown function of local variables. We note that 
the transition rate improves in accuracy up to $r\simeq L/2$ as $r$ increases. 
For each $r$, in the same way as the application to the ASEP, 
$C^h((n_{1\pm j})_{j=1}^{r})$ is iteratively determined as  
$C^{(l+1)\delta h}((n_{1\pm j})_{j=1}^{r}) =C^{l \delta h}((n_{1\pm j})_{j=1}^{r}) \left \langle e^{\tau \delta h A(\omega)}
\right \rangle_{F_1 \bv n}^{l\delta h} / \left \langle e^{\tau \delta h A(\omega)}
\right \rangle_{\bv n}^{l\delta h}$, where $\bv n = (0,n_2,...,n_{r+1}, 0, ... , 0, n_{L-r+1},... n_{L} )$.
First, for small $L$, we check the validity of the obtained $C^h((n_{1\pm j})_{j=1}^{r})$ by comparing it with the result obtained from the diagonalization of the matrix (\ref{matrixL}). An example of the result is shown in Fig. \ref{fig_modification_of_fluctuation_theorem} (i). 
Next, we fix relatively large $L$, and we obtain $G(h)$ for several values of $r$. 
We plot the obtained graph of $G(h)$ in 
Fig. \ref{fig_modification_of_fluctuation_theorem} (ii). In the same figure, 
we also plot the result obtained by employing the population dynamics method \cite{Populationdynamics,Populationdynamics2}.  
We observe that the curve for $r=4$ appears to be sufficient to explain the kinklike behavior of $G(h)$ near $h=0$.
This result suggests that the long-range interactions for the modified 
transition rate $w^h_{\rm eff}(\bv n\rightarrow  \bv n')$  may not be relevant 
to the dynamical phase transition observed in $G(h)$. 
The long-range nature of the effective interactions has also been studied very recently in
Ref. \cite{JackRecent} for the  East model. 
In order to investigate the singular behavior of $G(h)$ in greater detail,
a scaled biasing parameter $\tilde h \equiv h L$ has been used in this system \cite{bodineau1,bodineau2}. 
It has been mathematically proved that $\tilde G(\tilde h)=G(\tilde h /L)$
is not an analytic function in the limit $L \to \infty$ \cite{bodineau1}; however, 
the nature of the singularity has not thus far been elucidated. 
In \cite{suppl3}, we show that our method can also be applied to obtain the reliable $L$ dependence of $\tilde G(\tilde h)$. 


{\it Conclusion.}---
In this Letter, we formulated the evolution rule (\ref{whmesured})
in an exponential family for large deviation statistics. By this 
method, rare events are identified as typical events in modified 
systems, which are continuously generated via an iterative measurement-and-feedback procedure. 
For spatially extended many-body systems, where the 
number of degrees of freedom increases as an exponential function
of the system size, we also proposed a method for obtaining 
an {\it effective description of the exponential family} as a natural 
extension of our method. As examples of application of our method,
we studied an ASEP and a FA model. By numerical experiments with our method, 
we observed that the exponential family of the ASEP was well described by
another reparametrized ASEP, and
that the kinklike behavior of $G(h)$ of the FA model near $h=0$ was shown in an effective description without long-range interactions.
By performing further systematic 
numerical studies, we expect to obtain more quantitative information 
for large deviation statistics in spatially extended many-body systems. 
We believe that the formula  (\ref{mathBasis}) plays a fundamental 
role in large deviation theory, 
and, furthermore, we believe that our method provides a practically useful algorithm for
numerical experiments of large deviation statistics.

T. N. thanks F. van Wijland and V. Lecomte
for related discussions.
This study was supported by Grant-in-Aid for JSPS Fellows No. 247538,
KAKENHI No. 2234019, No. 25103002, and the JSPS Core-to-Core program ``Nonequilibrium Dynamics of Soft Matter and Information.''


\clearpage

\begin{center}
{\bf Supplemental material for ``Computation of large deviation statistics via iterative
measurement-and-feedback procedure''}
\end{center}

\noindent
{\bf 1. Derivation of the theoretical basis (5) in the text:} 

\vspace{2mm}

Here, we derive (5) in the text.
As a preliminary, we consider a matrix
\begin{equation}
L^{h,0}_{\bv n,\bv n^{\prime}}\equiv w(\bv n^{\prime}\rightarrow \bv n) e^{h\alpha(\bv n^{\prime}
\rightarrow \bv n)} - \lambda (\bv n) \delta_{\bv n, \bv n^{\prime}}.
\label{matrixL_suppli}
\end{equation}
Let $\phi^{h,0}$ and $G^{h,0}$ be the left eigenvector and the eigenvalue
of the largest eigenvalue problem of (\ref{matrixL_suppli}). 
Then, we define a modified transition rate as
\begin{equation}
u^{h}(\bv n^{\prime}\rightarrow \bv n) \equiv w(\bv n^{\prime}\rightarrow \bv n) e^{h \alpha(\bv n^{\prime} \rightarrow \bv n)} \phi^{h,0}(\bv n)/\phi^{h,0}(\bv n^{\prime}).
\end{equation}
The path probability density in the steady state generated by
$u^h$ is connected with the exponential family. We denote by $\left \langle f(\omega) \right \rangle ^{u^h}$ the expected value of a time-extensive quantity $f(\omega)$ with respect to the path probability density. Then, 
it has been known that
\begin{equation}
\left \langle f(\omega) \right \rangle^{u^h}
\simeq \frac{\left \langle f(\omega) e^{h\tau A(\omega)} \right \rangle }{ \left \langle e^{h\tau A(\omega)} \right \rangle}.
\label{suppl_propertyofu}
\end{equation}
See Ref. \cite{JackSollich_Suppl} for the details of the derivation.
This relation has been re-invented and used in many works, for example, in Refs.
\cite{JackSollich_Suppl,Discussion_of_modifiedTransitionrate_Suppl,Chetrite_Touchette_Suppl}.
With the aid of (\ref{suppl_propertyofu}), 
the relation (5) in the text may be derived from the equivalence between the transition rate $u^h$ and
the transition rate $w^h$ introduced in the text.


In order to show the equivalence, we first consider the matrix defined by $u^h$:
\begin{equation}
L^{h,h^{\prime}}_{\bv n,\bv n^{\prime}}\equiv u^{h^{\prime}}(\bv n^{\prime}\rightarrow \bv n) e^{h\alpha(\bv n^{\prime}
\rightarrow \bv n)} - \lambda^{u^{h^{\prime}}}(\bv n) \delta_{\bv n, \bv n^{\prime}}
\label{Suppl_matrix2}
\end{equation}
with $\lambda^{u^{h^{\prime}}}(\bv n)\equiv \sum_{\bv n^{\prime}} u^{h}(\bv n\rightarrow \bv n^{\prime})$.
We denote the left eigenvector and the eigenvalue of the largest eigenvalue problem of (\ref{Suppl_matrix2}) by $\phi^{h,h^{\prime}}$ and $G^{h,h^{\prime}}$.
We then present 
{\it the multiplicative property} for the eigenvector and {\it the additive property}
for the eigenvalue of the largest eigenvalue problem of (\ref{Suppl_matrix2}), which are
\begin{equation}
\phi^{h+h^{\prime},0}= \phi^{h,h^{\prime}} \phi^{h^{\prime},0}
\label{Suppl_multiplicative}
\end{equation}
and
\begin{equation}
G^{h+h^{\prime},0}=G^{h,h^{\prime}} + G^{h^{\prime},0}.
\label{Suppl_additive}
\end{equation}
The proof is the following.
First, we write the eigenvalue equations for $\phi^{h+h^{\prime},0}$, $\phi^{h^{\prime},0}$, and $\phi^{h,h^{\prime}}$ as
\begin{equation}
\sum_{\bv n^{\prime}}w(\bv n\rightarrow \bv n^{\prime}) e^{(h+h^{\prime})\alpha(\bv n
\rightarrow \bv n^{\prime})}\frac{\phi^{h+h^{\prime},0}(\bv n^{\prime})}{\phi^{h+h^{\prime},0}(\bv n)} - \lambda (\bv n) =G^{h+h^{\prime},0},
\label{Suppl_Eigen1}
\end{equation}
\begin{equation}
\sum_{\bv n^{\prime}}w(\bv n\rightarrow \bv n^{\prime}) e^{h^{\prime}\alpha(\bv n
\rightarrow \bv n^{\prime})}\frac{\phi^{h^{\prime},0}(\bv n^{\prime})}{\phi^{h^{\prime},0}(\bv n)} - \lambda (\bv n) =G^{h^{\prime},0},
\label{Suppl_Eigen2}
\end{equation}
and
\begin{equation}
\begin{split}
&\sum_{\bv n^{\prime}}w(\bv n\rightarrow \bv n^{\prime}) e^{(h+h^{\prime})\alpha(\bv n
\rightarrow \bv n^{\prime})}\frac{\phi^{h,h^{\prime}}(\bv n^{\prime}) \phi^{h^{\prime},0}(\bv n^{\prime})}{\phi^{h,h^{\prime}}(\bv n) \phi^{h^{\prime},0}(\bv n)} - \lambda^{u^{h^{\prime}}} (\bv n)  \\
&=G^{h,h^{\prime}}.
\end{split}
\label{Suppl_Eigen3}
\end{equation}
Since the first term of (\ref{Suppl_Eigen2}) is equal to the second term of (\ref{Suppl_Eigen3}), these terms cancel each other when we sum (\ref{Suppl_Eigen2}) and (\ref{Suppl_Eigen3}). We thus obtain
\begin{equation}
\begin{split}
&\sum_{\bv n^{\prime}}w(\bv n\rightarrow \bv n^{\prime}) e^{(h+h^{\prime})\alpha(\bv n
\rightarrow \bv n^{\prime})}\frac{\phi^{h,h^{\prime}}(\bv n^{\prime}) \phi^{h^{\prime},0}(\bv n^{\prime})}{\phi^{h,h^{\prime}}(\bv n) \phi^{h^{\prime},0}(\bv n)} - \lambda (\bv n)  \\
&=G^{h,h^{\prime}} + G^{h^{\prime},0}.
\end{split}
\label{Suppl_Eigen4}
\end{equation}
From the Perron-Frobenius theory for irreducible matrices \cite{Seneta_Suppl}, the positive eigenvector of $L_{\bv n,\bv n^{\prime}}^{h,0}$ is unique. Thus, by comparing 
(\ref{Suppl_Eigen1}) with (\ref{Suppl_Eigen4}), we arrive at 
(\ref{Suppl_multiplicative}) and (\ref{Suppl_additive}).


Second, we show a relation between the eigenvector $\phi^{h,h^{\prime}}$ and $\left \langle e^{\tau  h A(\omega)} \right \rangle_{\bv n}$.
We start with the time-evolution equation of
$\left \langle e^{h \tau A(\omega)} \delta_{\bv n(\tau), \bv n}\right \rangle_{\bv n_0}$,
\begin{equation}
\frac{\partial \left \langle e^{h \tau A(\omega)} \delta_{\bv n(\tau), \bv n}\right \rangle_{\bv n_0}}{\partial \tau}=\sum_{\bv n^{\prime}}L_{\bv n,\bv n^{\prime}}^{h,0}  \left \langle e^{h \tau A(\omega)} \delta_{\bv n(\tau), \bv n^{\prime}}\right \rangle_{\bv n_0}.
\label{supple_timeevolution}
\end{equation}
See Ref. \cite{garrahan2_Suppl} for the derivation. 
We can rewrite this expression as
\begin{equation}
\frac{\partial \left \langle e^{h \tau A(\omega)} \right \rangle_{\bv n}}{\partial \tau}=\sum_{\bv n^{\prime}}L_{\bv n^{\prime},\bv n}^{h,0}  \left \langle e^{h \tau A(\omega)} \right \rangle_{\bv n^{\prime}}.
\end{equation}
This time evolution equation indicates an asymptotic expression
\begin{equation}
\phi^{h, 0}(\bv n) \simeq \left \langle e^{\tau  h A(\omega)} \right \rangle_{\bv n}.
\label{suppl_phi_original}
\end{equation}
We note that the same method may be applied to the system with $u^h$. In this case, the matrix appeared in (\ref{supple_timeevolution}) is replaced by $L^{h,h^{\prime}}_{\bv n,\bv n^{\prime}}$, which leads to
\begin{equation}
\phi^{h, h^{\prime}}(\bv n) \simeq \left \langle e^{\tau  h A(\omega)} \right \rangle_{\bv n}^{u^{h^{\prime}}}.
\label{suppl_phi_original2}
\end{equation}

Finally, from the results above, we show the equivalence between $w^h$ and $u^h$.
We fix an increment $\delta h$.
First, $w^{\delta h}$ is defined as
\begin{equation}
w^{\delta h}(\bv n\rightarrow \bv n^{\prime}) = w(\bv n\rightarrow \bv n^{\prime}) e^{\delta h \alpha(\bv n \rightarrow \bv n^{\prime})}\frac{\left \langle e^{ \tau \delta h A(\omega)} \right \rangle_{\bv n^{\prime}}}{\left \langle e^{\tau \delta h A(\omega)} \right \rangle_{\bv n}}.
\end{equation}
With the relation (\ref{suppl_phi_original}), the definition leads to
\begin{equation}
w^{\delta h}\simeq u^{\delta h}.
\label{suppl_re1}
\end{equation}
The next transition rate $w^{2\delta h}$ is defined as
\begin{equation}
\begin{split}
&w^{2\delta h}(\bv n\rightarrow \bv n^{\prime}) = w^{\delta h}(\bv n\rightarrow \bv n^{\prime}) e^{\delta h \alpha(\bv n \rightarrow \bv n^{\prime})}\frac{\left \langle e^{ \tau \delta h A(\omega)} \right \rangle_{\bv n^{\prime}}^{\delta h}}{\left \langle e^{ \tau \delta h A(\omega)} \right \rangle_{\bv n}^{\delta h}},
\end{split}
\end{equation}
where $\left \langle \ \right \rangle_{\bv n}^{\delta h}$ is the expected value in the modified system, which is equivalent to $\left \langle \ \right \rangle_{\bv n}^{u^{\delta h}}$ due to (\ref{suppl_re1}).
Then, from the multiplicative property (\ref{Suppl_multiplicative}) with (\ref{suppl_phi_original2}),
we obtain
\begin{equation}
w^{2\delta h}\simeq u^{2\delta h}.
\label{suppl_re2}
\end{equation}
By iterating this procedure, we thus arrive at the
equivalence between $u^{h}$ and $w^{h}$. That is,
a set of transition rates
\begin{equation}
\begin{split}
&w^{l \delta h}(\bv n\rightarrow \bv n^{\prime}) = w(\bv n\rightarrow \bv n^{\prime}) e^{l \delta h \alpha(\bv n \rightarrow \bv n^{\prime})}\prod_{k=0}^{l-1}\frac{\left \langle e^{ \tau \delta h A(\omega)} \right \rangle_{\bv n^{\prime}}^{k \delta h}}{\left \langle e^{ \tau \delta h A(\omega)} \right \rangle_{\bv n}^{k \delta h}}
\end{split}
\end{equation}
with $l=0,1,2, ...$ satisfies
\begin{equation}
w^{l\delta h}\simeq u^{l\delta h}
\end{equation}
for $l=0,1,2, ...$.

\vspace{5mm}

\noindent
{\bf  2. Scaled cumulant generating function for a FA model:} 

\vspace{2mm}

\noindent
{\bf (i). Scaled cumulant generating function and obtained result:}

In order to investigate the singular behavior of $G(h)$ in greater detail,
a scaled biasing parameter $\tilde h \equiv h L$ has been used in one-dimensional FA model \cite{bodineau1_Suppl,bodineau2_Suppl}. 
It has been mathematically proved that $\tilde G(\tilde h)=G(\tilde h/L)$
is not an analytic function in the limit $L \to \infty$ \cite{bodineau1_Suppl}; however, 
the nature of the singularity has not thus far been elucidated. 
The problem has been numerically studied by employing the population dynamics method 
\cite{bodineau2_Suppl}. However, it seems that the result does not 
exhibit good convergence of $\tilde G(\tilde h)$ for relatively large values of $L$. 
In this supplemental material, we show that our method can be applied to obtain the reliable $L$ dependence of $\tilde G(\tilde h)$ up to $L=60$.

\begin{figure}[tbh]
\includegraphics[width=4cm]{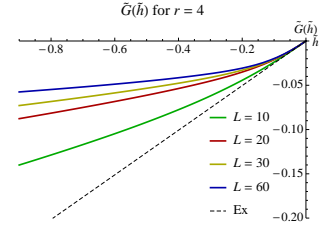}
\caption{(Color online) 
Cumulant generating functions obtained from 
our computational method for a FA model with $c=0.3$. 
We plot $\tilde G(\tilde h)\equiv G(\tilde h/L)$ for various values of $L$ with $r=4$ fixed. We also plot the straight line (Ex) of the slope $4c^2(1-c)$, which is the expected value of the activity in the original system with $h=0$.
} 
\label{figsupp0}
\end{figure}

In the text, for the one-dimensional FA model, we introduced the effective transition rate defined as
\begin{equation}
\begin{split}
w^h_{\rm eff}(\bv n\rightarrow  F_i \bv n) \equiv w(\bv n\rightarrow  F_i \bv n)
e^{h}\left [C^h((n_{i\pm j})_{j=1}^{r}) \right ]^{1-2n_i},
\end{split}
\end{equation}
where $C^h((n_{i\pm j})_{j=1}^{r})$ was an unknown function of local variables characterized by the truncating number $r$. 
By investigating $\tilde G(\tilde h)$ via the effective description
with several truncating numbers $r$ and system sizes $L$, we will judge that the result with 
$r=4$ is sufficiently accurate to obtain the true $L$-dependence of $\tilde G(\tilde h)$. 
We present this argument in the next section.
Here, we show the obtained $L$ dependence of $\tilde G(\tilde h)$. We plot $\tilde G(\tilde h)$ with $r=4$ fixed for various values of $L$ in Fig. \ref{figsupp0}. 
Although it has been conjectured that 
$\tilde G(\tilde h)$ contains a non-differentiable point 
in the limit $L \to \infty$ \cite{bodineau1_Suppl}, our result 
does not show any clear sign of such a point in $\tilde G(\tilde h)$ 
up to $L=60$.  

\vspace{2mm}

\noindent
{\bf (ii). Truncating number of a FA model:}

In this section, we present the
evidence that 
$r=4$ is sufficiently large to describe the true $L$-dependence of the 
scaled function 
$\tilde G(\tilde h)\equiv G(L^{-1} \tilde h)$.


\begin{figure}[tbh]
\includegraphics[width=4cm]{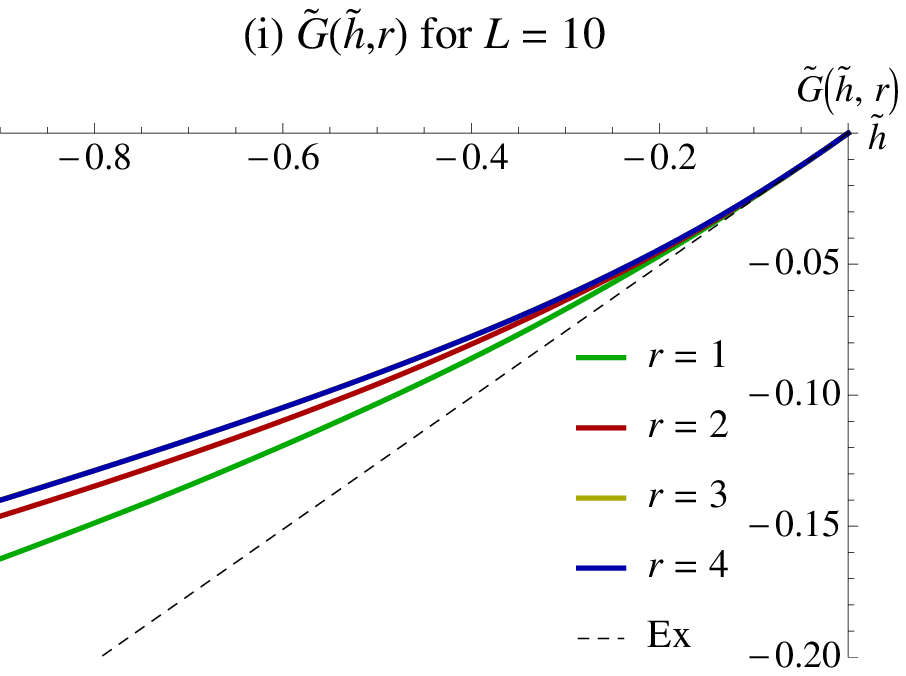}
\includegraphics[width=4cm]{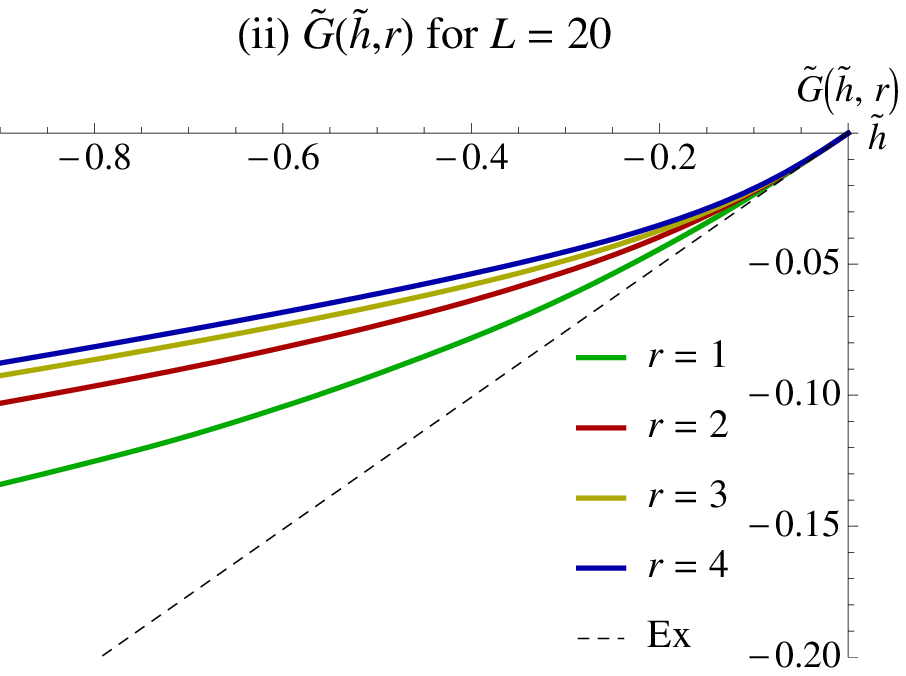}
\includegraphics[width=4cm]{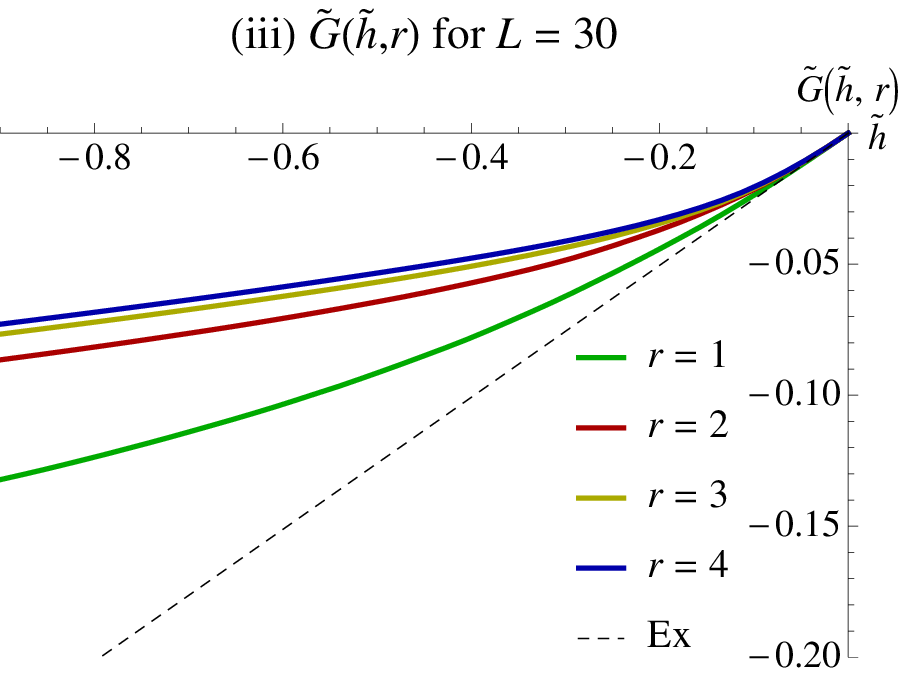}
\includegraphics[width=4cm]{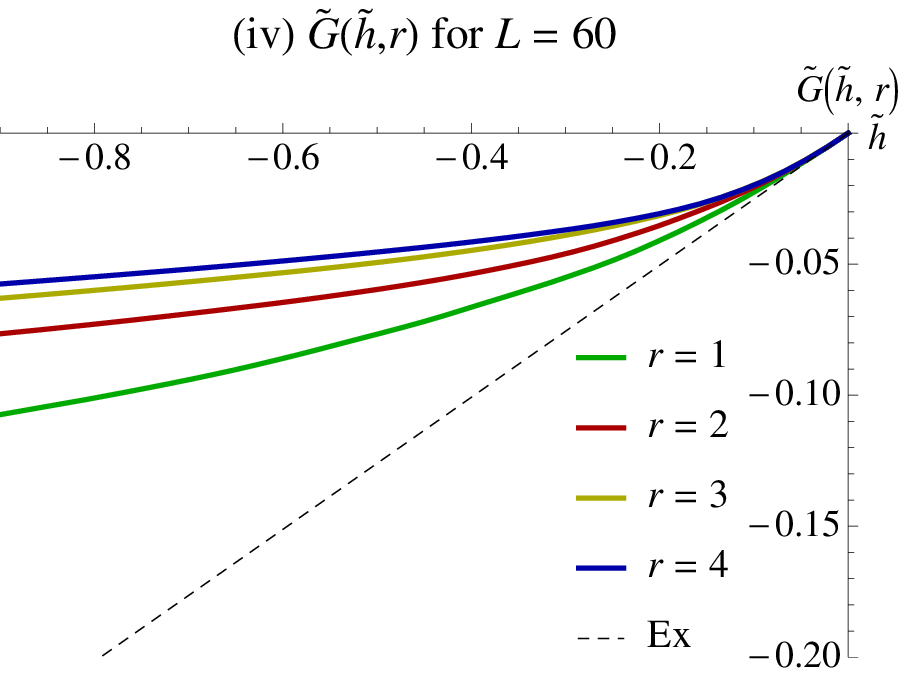}
\caption{(Color online) 
$\tilde G(\tilde h,r)$ for $r=1,2,3$, and $4$ with $L=10$ (i), 20 (ii), 30 (iii), and 60 (iv). We set $c=0.3$. We also plot the straight line (Ex) with 
the slope $4c^2(1-c)$ in each figure.
} 
\label{figuresupp1}
\end{figure}

First, we denote by $\tilde G(\tilde h,r)$ our calculation result of $\tilde G(\tilde h)$ obtained with truncating number $r$.
In Fig. \ref{figuresupp1}, we show  
$\tilde G(\tilde h,1)$, $\tilde G(\tilde h,2)$, $\tilde G(\tilde h,3)$, and $\tilde G(\tilde h,4)$ for $L=10$ (i), $20$ (ii), $30$ 
(iii), and $60$ (iv). We also plot the straight line (Ex) with 
the slope $4c^2(1-c)$ in each figure, which is the expected 
value of the activity in the original system with $\tilde h=0$.
This straight line is equivalent to $\tilde G(\tilde h,0)$, since
there are no modifications for $r=0$. 
In Fig. \ref{figuresupp1}, we observe that 
the differences between $\tilde G(\tilde h)$ with $r=3$ and that 
with $r=4$ are small even for larger $L$ (say $L=20,30$, and $60$).


\begin{figure}[tbh]
\includegraphics[width=4cm]{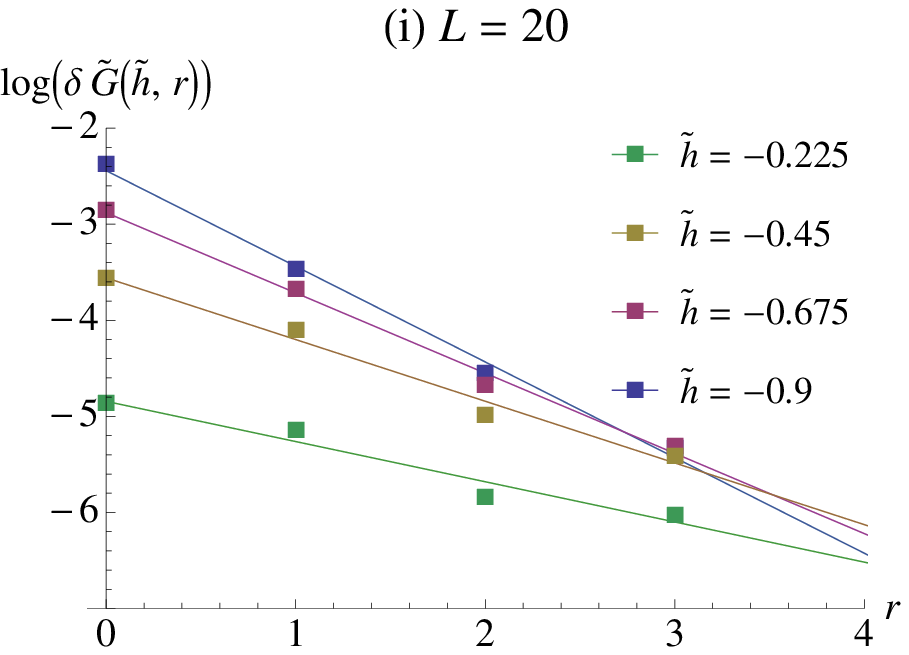}
\includegraphics[width=4cm]{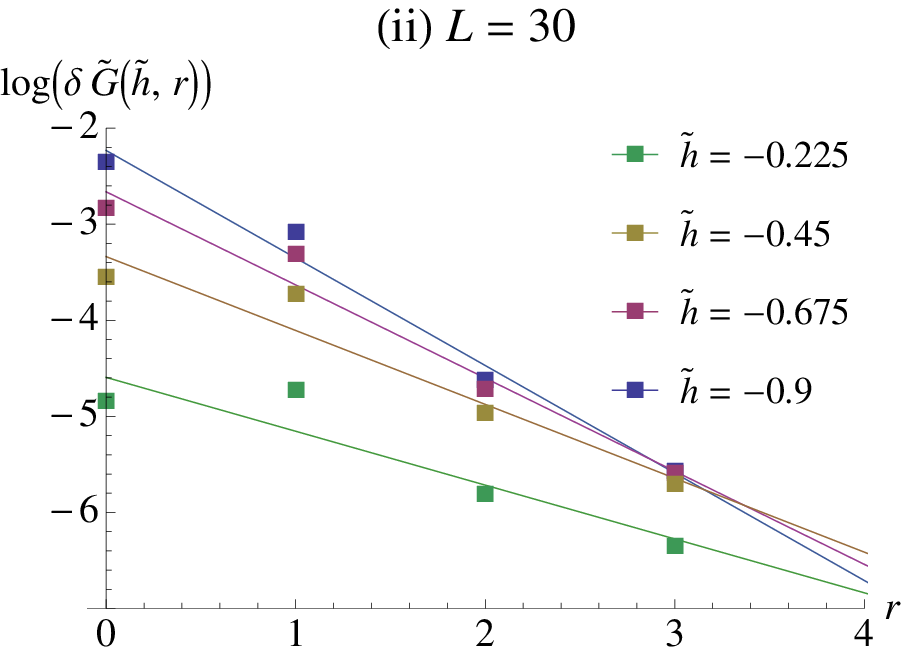}
\includegraphics[width=4cm]{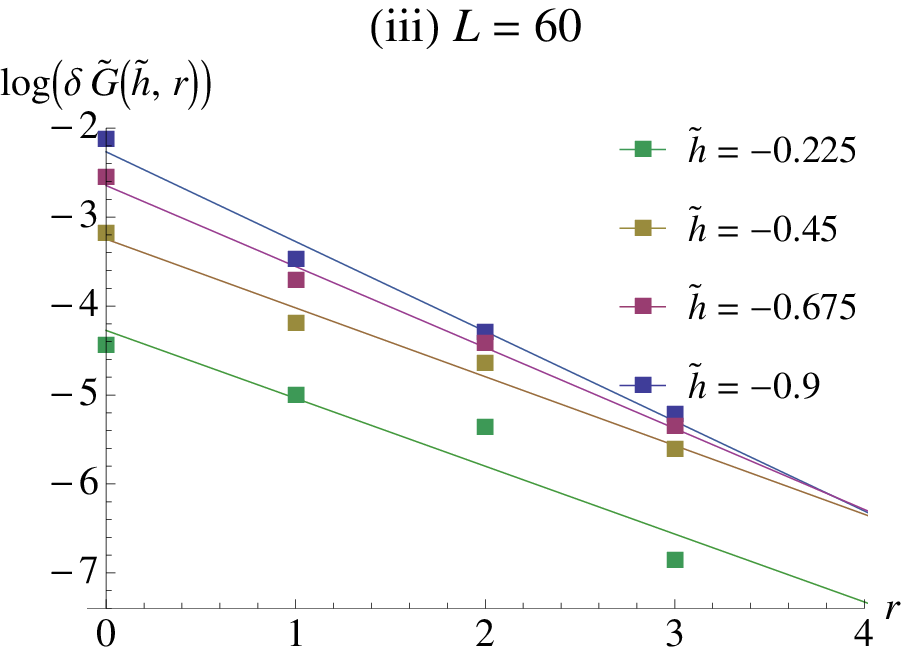}
\caption{(Color online) 
The logarithm of the difference function $\delta \tilde G(\tilde h ,r)$ for $r=0,1,2$, and
$3$ with $L=20$ (i), 30 (ii), and 60 (iii). We set $c=0.3$.
} 
\label{figuresupp2}
\end{figure}

We then quantify these small differences by introducing a difference function 
\begin{equation}
\delta \tilde G(\tilde h,r) = \tilde G(\tilde h,r+1) - \tilde G(\tilde h,r).
\end{equation}
In Fig. \ref{figuresupp2}, we plot the logarithm of $\delta G(r,h)$ as a function of $r$ for $\tilde h=-0.225,-0.45,-0.675$ and $-0.9$ with $L=20$ (i), $30$ (ii), and $60$ 
(iii). In the same figure, we also plot straight lines, which were obtained from a least squares fit of the data points.
We observe that the decay of $\delta \tilde G$ with increasing $\tilde r$ is exponentially fast (or faster than exponential decay).
We thus expect that larger $r$ is not needed to obtain the correct $\tilde G (\tilde h)$.


\begin{figure}[tbh]
\includegraphics[width=4cm]{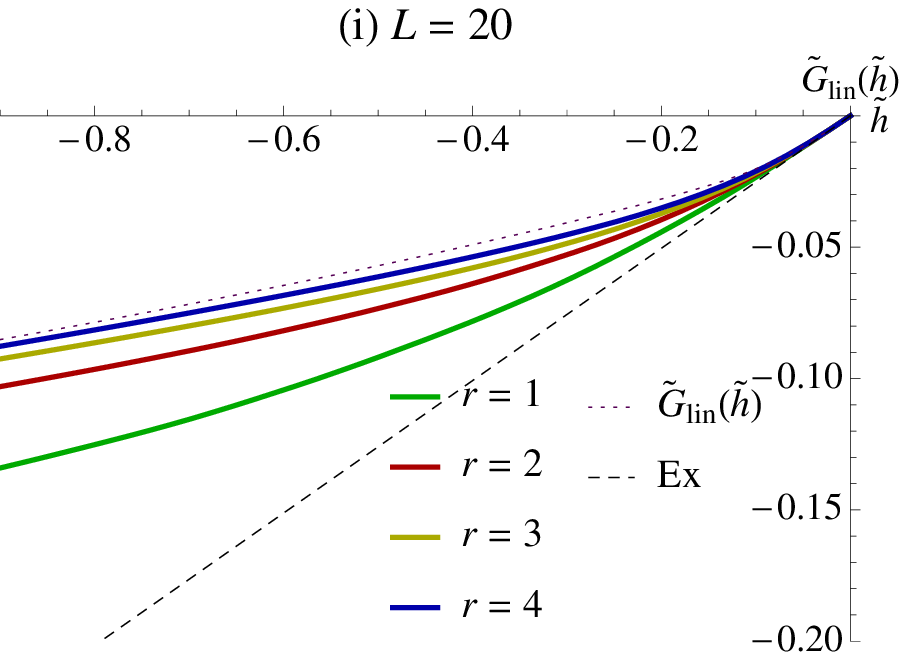}
\includegraphics[width=4cm]{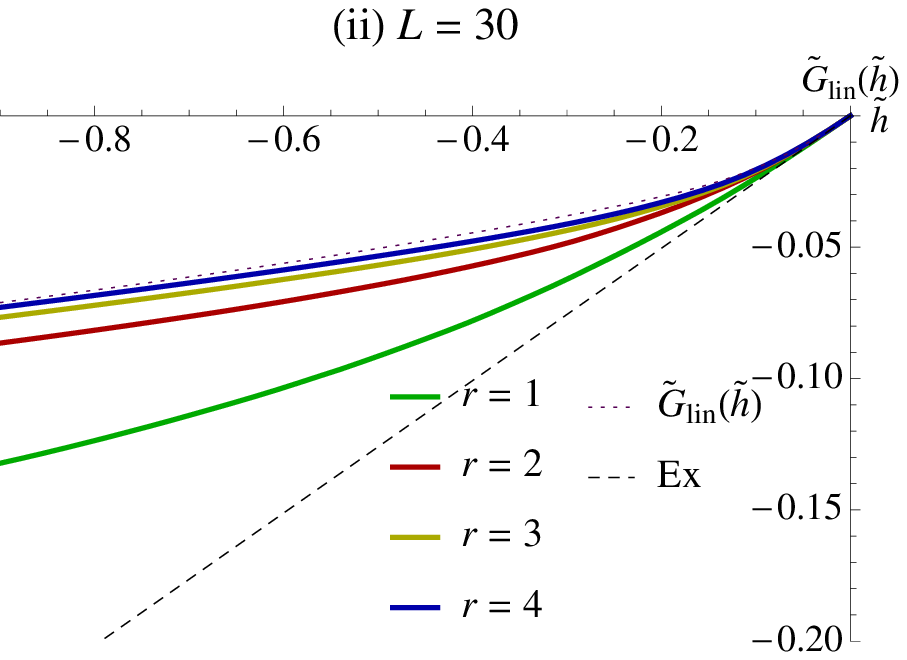}
\includegraphics[width=4cm]{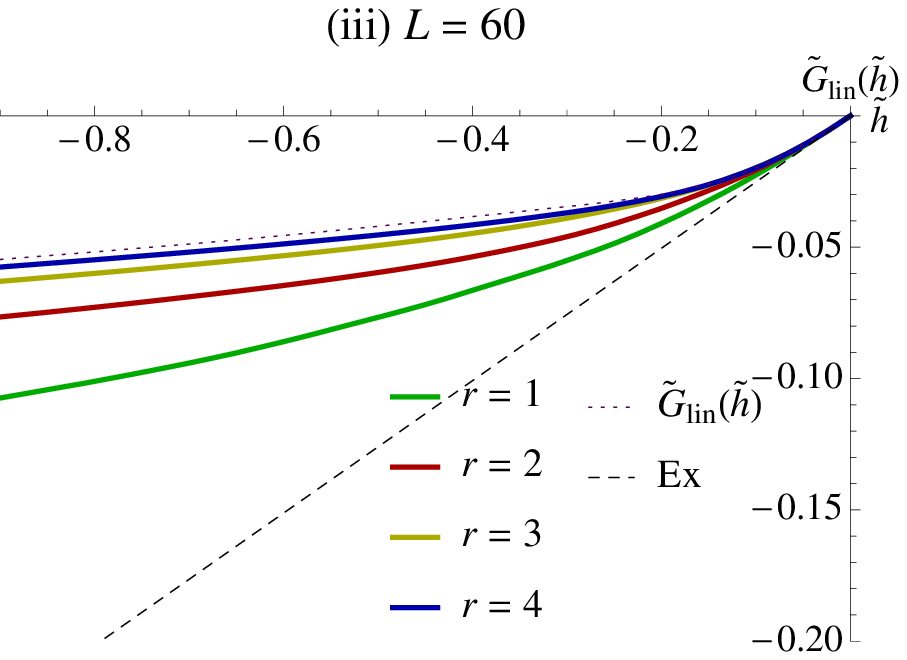}
\caption{(Color online) 
$\tilde G(\tilde h,r)$ and $\tilde G_{\rm lin}(\tilde h)$ for $r=1,2,3,$ and $4$ with $L=20$ (i), 30 (ii), and 60 (iii). We set $c=0.3$.
} 
\label{figuresupp3}
\end{figure}

Finally, by assuming the shape of $\delta \tilde G(\tilde h,r)$ as an exponentially decaying function, we estimate possible errors due to the truncation of $r$ with $r=4$.
Inspired by the result of Fig. \ref{figuresupp2}, we define
\begin{equation}
\delta \tilde G_{\rm lin}(\tilde h,r) = e^{a(\tilde h) r + b(\tilde h)},
\label{assumption_linear_Supp2}
\end{equation}
where $a(\tilde h)r + b(\tilde h)$ is a linear function of $r$, which is determined from
the least squares fit of data points $\log \delta \tilde G(\tilde h,r=0)$, $\log \delta \tilde G(\tilde h,r=1)$, $\log \delta \tilde G(\tilde h,r=2)$, and $\log \delta \tilde G(\tilde h,r=3)$. 
The examples of the linear function are solid lines in Fig. \ref{figuresupp2}.
Then, by using $\delta \tilde G_{\rm lin}$, we define an interpolation function $\tilde G_{\rm lin}(\tilde h,r)$ of $\tilde G(\tilde h,r)$ for larger $r$ as
\begin{equation}
\tilde G_{\rm lin}(\tilde h,r) \equiv \tilde G (\tilde h,4) + \sum_{s=4}^{r-1}
 \delta \tilde G_{\rm lin}(\tilde h,s)
\end{equation}
for $r=5,6,...$.
Since $\tilde G(\tilde h,r)$ with $r\simeq L/2$ is equal to
$\tilde G(\tilde h)$, we thus define an interpolation function of $\tilde G(\tilde h)$
as
\begin{equation}
\tilde G_{\rm lin}(\tilde h) \equiv \tilde G_{\rm lin}(\tilde h,L/2-1).
\end{equation}
We plot $\tilde G_{\rm lin}(\tilde h)$ with $\tilde G(\tilde h,r)$ for $r=1,2,3$, and 4 in Fig. \ref{figuresupp3}.
In the figure, the differences between $\tilde G_{\rm lin}(\tilde h)$ and $\tilde G(\tilde h,4)$ are quite small, so that we judge that $r=4$ is sufficiently large to describe the true $L$-dependence of $\tilde G(\tilde h)$.

\end{document}